\documentclass[showpacs,aps,preprint,showkeys]{revtex4}
\begin{document}
\title{On the Hamilton-Jacobi formalism for fermionic systems }
\author{C. Ram\'{\i}rez}
\email{cramirez@fcfm.buap.mx}
\altaffiliation[Permanent address:]{
Facultad de Ciencias F\'{\i}sico\\
\vskip -1truecm
Matem\'aticas, Universidad Aut\'onoma de Puebla,
P.O. Box 1364, 72000 Puebla, M\'exico.}
\affiliation{Instituto de F\'{\i}sica de la Universidad de Guanajuato,\\
P.O. Box E-143, 37150 Le\'on Gto., M\'exico}
\author{P. A. Ritto}
\email{parmunacar@yahoo.com.mx}
\altaffiliation[Permanent address:]{
Facultad de Ingenier\'{\i}a, Universidad\\
\vskip -1truecm
Aut\' onoma del Carmen,
24180 Cd. del Carmen, Camp., Mexico.}
\affiliation{Facultad de Ciencias F\'{\i}sico Matem\'aticas,\\
Universidad Aut\'onoma de Puebla,\\
P.O. Box 1364, 72000 Puebla, M\'exico.}
\date{\today}

\vskip -.5truecm

\begin{abstract}
The Hamilton-Jacobi formalism for fermionic systems is studied. We derive the HJ equations from
the canonical transformation procedure, taking into account the second class constraints typical
of these systems. It is shown that these constraints ensure the consistency of the solution, according to the characteristics of fermionic systems.
The explicit solutions for simple examples are computed.
Some aspects related to canonical transformations and to quantization are discussed.
\end{abstract}
\maketitle

\section*{Introduction}
The basic property of fermions, half-integer spin, is a property
of microscopic particles and is described by Quantum Mechanics.
However the classical mechanics of fermions has been a very useful
tool in their study, in particular regarding their path integral
formulation \cite{1}. Their Lagrangian and Hamiltonian
formulations are well known, in particular the Hamiltonian requires of the Dirac formalism of singular systems
\cite{2,teitelboim}, as far as they have second class constraints. The understanding of the Hamilton Jacobi
(HJ) formalism
can be helpful to understand better quantum systems. It has been applied to singular systems, as is the case of general
relativity in the case of the Wheeler-DeWitt equations. In this
case the HJ formulation amounts to set the constraints of the
Hamiltonian formalism as differential equations on the wave
function \cite{rovelli}. For the WKB approximation of singular systems, in
particular fermionic ones \cite{zumino}, it would be also useful
to have a systematic way to obtain the HJ formulation. Here an
effort is made with the aim to understand this issue better.
Although fermionic systems have been widely studied, their HJ
formulation has been less studied. It has been worked out for
bosonic constrained systems \cite{guler}, where
the constraints, obtained from the Hamiltonian formulation, are
written as separated equations. For fermionic systems, a technique
to handle a particular problem in the HJ formalism was proposed in
\cite{zumino}. However, the most general situation is not
discussed. In \cite{brasilenios} the G\"uler formalism \cite{guler} is generalized to
include fermionic variables, by the application of the usual
concepts of classical analysis, whose properties, nevertheless, are not of general validity when dealing
with nilpotent quantities. In this formulation all the constraints of the Hamiltonian formulation are kept as additional equations to the actual HJ equation. In \cite{tabunschyk} a formulation is given,
which strongly relies on the Hamiltonian formulation.

Here we give a formulation of the HJ equation for fermionic systems, which
is obtained as usual for bosonic theories, from the variation of the action in canonical
coordinates, considering the transformation to constant new
coordinates \cite{pavel}. As in the G\"{u}ler formalism, we apply the ``second class'' constraints characteristic of fermionic theories, as additional equations.
We show that these equations have two important consistency consequences. First, from the way we obtain the HJ equation, there are two integration constants for each fermionic degree of freedom, and we get a set of equations among these constants, which reduce their number to half, as it must be for a first order theory. Further, related to this last fact, as noted in \cite{heno}, boundary conditions have to be added to fermionic actions. This means that also the generator functions of canonical transformations must satisfy boundary conditions. It is shown that the mentioned equations ensure that these boundary conditions are automatically satisfied.

In order to verify the validity of the resulting equations, we consider two examples of simple
fermionic systems. The solutions to these equations are found to be the same as the
solutions of the Euler-Lagrange equations.

In the first section the Lagrangian and Hamiltonian formalisms
for bosonic and fermionic systems are reviewed. In the second section the HJ
equation for fermionic systems is given. In the next two sections illustrative fermionic systems of one and
two variables are considered. In the framework of the second example, in the next section, the relation to the canonical transformations is established in a more precise way and, in the last section, the relation to the quantum theory is discussed.

\section*{Fermionic Mechanics}
Let us consider a classical system described by $n$ bosonic, even Grassmann,
degrees of freedom $q=(q_1,q_2,...,q_n)$ and $\mu$ fermionic, odd
Grassmann, degrees of freedom
$\psi=(\psi_1,\psi_2,...,\psi_\mu)$. These variables obey the
relations
\begin{eqnarray}
q_iq_j-q_jq_i&=&0\qquad i,j=1,2,...,n\quad,\nonumber\\
q_i\psi_\alpha-\psi_\alpha q_i&=&0\quad,\label{grasman}\\
\psi_\alpha\psi_\beta+\psi_\beta\psi_\alpha&=&0\qquad \alpha,
\beta=1,2,...,\mu\quad.\nonumber
\end{eqnarray}
In this case, the Lagrangian function depends on the $q$'s, on the
$\psi$'s, and on their respective time derivatives
\begin{equation}
L=L(q,\psi,\dot q,\dot\psi,t)= L(Q,\dot Q,t),\label{lagfer}
\end{equation}
where $Q= (q,\psi)$.

If we variate the corresponding action
\begin{equation}
\delta S=\int_{t_1}^{t_2}\left (\delta q_{i}\frac{\partial L}{\partial q_i}+\delta
\dot{q}_{i}\frac{\partial L}{\partial\dot{q}_i}+\delta
\psi_{\alpha}\frac{\partial L}{\partial \psi_\alpha}+\delta
\dot{\psi}_{\alpha}\frac{\partial L}{\partial \dot{\psi}_\alpha}\right ),
\end{equation}
then, in order to get the Euler-Lagrange equations,
\begin{equation}
\frac{d}{dt}\left (\frac{\partial L}{\partial \dot{Q}_k}\right
)-\frac{\partial L}{\partial Q_k}=0\qquad k=1,2,...,n+\mu,\label{ecsfer}
\end{equation}
suitable boundary conditions must be imposed. For the bosonic variables it can be done as usual by fixing each of them at both extrema. However, for each of the fermionic degrees of freedom, as far as they are first order in the velocities, only one boundary condition can be fixed. In this case, for consistency, suitable boundary terms must be added to the action \cite{heno}. For example, if the fermionic kinetic term is
$L_{kin}=\frac{i}{2}g^{\alpha\beta}\psi_\alpha\dot\psi_\beta$, then the corrected action is given by
\cite{heno},
\begin{equation}
S- \frac{i}{2}g^{\alpha\beta}\psi_\alpha(t_1)\psi_\beta(t_2),\label{boundary}
\end{equation}
with the boundary conditions $\delta[\psi_\alpha(t_1)+ \psi_\alpha(t_2)]=0$, that is
$\psi_\alpha(t_2)=-\psi_\alpha(t_1)+\xi_\alpha$, where $\xi_\alpha$ are constant anticommuting quantities.

The Hamiltonian is given by
\begin{equation}
H(Q,P,t)=\dot q p+\dot\psi\pi-L\equiv \dot QP-L\label{def}
\end{equation}
were $p_i\equiv{\partial L} /{\partial\dot{q}^i}$,
$\pi^\alpha\equiv{\partial L}/{\partial\dot{\psi}_\alpha}$ and $P\equiv(p,\pi)$.

The Lagrangian (\ref{lagfer}) is first order in the fermionic variables,
i.e. the kinetic term is
linear in fermionic velocities and the potential does not depend on
them. Therefore there are primary constraints,
\begin{equation}
\phi^\alpha=\pi^\alpha-f^\alpha(q,\psi),\label{constricciones}
\end{equation}
where $f^\alpha(q,\psi)$ are odd Grassmann functions. In the Dirac
formalism for constrained systems, these constraints turn out to be second class.
Further we suppose that there are no more constraints. Thus, due to the fact that
in the Hamiltonian (\ref{def}) the term
$\dot\psi_\alpha\pi^\alpha$ compensates the corresponding kinetic term
in the Lagrangian,  the canonical
Hamiltonian does not depend on the fermionic momenta,
\begin{equation}
H= H(Q,p,t),
\end{equation}
Therefore, if the Lagrangian is purely fermionic,
the Hamiltonian will be given by the potential.

\subsection*{HJ formalism for Grassmann variables}
In order to find the HJ equation for the preceding system, we
consider the variation of the action
\begin{equation}
S=\int_{t_1}^{t_2} L(Q,\dot{Q},t)dt+BT= \int_{t_1}^{t_2} \left
[\dot{Q}P-H(Q,p,t)\right ]dt+BT,
\end{equation}
where the second class constraints (\ref{constricciones}) and, as mentioned in the preceding section, suitable
fermionic boundary conditions ($BT$) added to the action, insure us that the variation
of the right hand side gives the correct equations of motion. For instance, if we consider the action
$L=\frac{i}{2}g^{\alpha\beta}\psi_\alpha\dot\psi_\beta-V(\psi)$, then we have
$\pi^\alpha=-\frac{i}{2}g^{\alpha\beta}\psi_\beta$ and if we impose the boundary conditions
$\delta[\psi_\alpha(t_1)+\psi_\alpha(t_2)]=0$, then,
\begin{eqnarray}
\delta S&=&\delta\left\{\int_{t_1}^{t_2}\left[\dot\psi_\alpha\pi^\alpha-H(\psi,t)\right]dt+\frac{i}{2}
\psi_\alpha(t_1)\pi^\alpha(t_2)\right\}\nonumber\\
&=&\int_{t_1}^{t_2}
\left(-\delta\psi_\alpha\dot\pi^\alpha+\dot\psi_\alpha\delta\pi^\alpha-
\delta\psi_\alpha\frac{\partial H}{\partial\psi_\alpha}\right)=
\int_{t_1}^{t_2}\delta\psi_\alpha\left(ig^{\alpha\beta}\dot\psi_\beta-
\frac{\partial V}{\partial\psi_\alpha}\right)=0.
\end{eqnarray}

Thus, the physical phase space is ($2n+\mu$)-dimensional hyperplane $\cal P$,
solution of (\ref{constricciones}).

Let us do a canonical transformation of coordinates, $(Q,P)\rightarrow(\tilde Q,\tilde P)$. In this case the
constraints (\ref{constricciones}) will transform to some constraints
\begin{equation}
\tilde\phi_\alpha(\tilde Q,\tilde P)=0.\label{constr}
\end{equation}
To obtain the HJ equation a variation of these actions is done,
\begin{eqnarray}
\delta S&=&\delta \left\{\int_{t_1}^{t_2}\left [\dot{Q}P-H(Q,p,t)\right ]dt +BT\right\}=0,\\
\delta S^\prime&=&\delta \left\{\int_{t_1}^{t_2}
\left[\dot{\tilde{Q}}\tilde{P}-H^\prime(\tilde{Q},\tilde{P},t)\right]dt+(BT)'\right\}=0,
\end{eqnarray}
The relation between integrands is,
\begin{equation}
\dot{Q}P-H(Q,p,t)=\dot{\tilde{Q}}\tilde{P}-H^\prime(\tilde{Q},\tilde{P},t)+\frac{dF}{dt}+K,
\end{equation}
where $K=\frac{(BT)'-BT}{t_2-t_1}$ and $F$ is a function, whose dependence on the phase space
coordinates and on time must be such that its variation at the boundary satisfies $\delta [F(t_2)-F(t_1)]=0$.

If now $F= F(Q,\tilde{P},t)-\tilde{Q}\tilde{P}$, then
\begin{equation}
\frac{dF}{dt}=\dot{Q}\frac{\partial F}{\partial
Q}+\dot{\tilde{P}}\frac{\partial F}{\partial\tilde{P}}+\frac{\partial
F}{\partial t}-\frac{d}{dt}(\tilde{Q}\tilde{P}) ,
\end{equation}
hence
\begin{equation}
\dot{Q}P-H(Q,p,t)=\dot{\tilde{Q}}\tilde{P}-H^\prime(\tilde{Q},\tilde{p},t)+\dot{Q}\frac{\partial
F}{\partial Q}+\dot{\tilde{P}}\frac{\partial
F}{\partial\tilde{P}}+\frac{\partial F}{\partial t}-\frac{d}{dt}(\tilde{Q}\tilde{P})+K .
\end{equation}

A factorization of this gives
\begin{equation}
\dot{Q}\left (P-\frac{\partial F}{\partial Q}\right )+\dot{\tilde{P}}\left
((-1)^{a_pa_q}\tilde{Q}-\frac{\partial F}{\partial\tilde{P}}\right )-\left
(H+\frac{\partial F}{\partial t}-H^\prime+K\right)=0,
\end{equation}
and additionally the constraints (\ref{constricciones}) and (\ref{constr}). The sign in the middle term corresponds to the interchange of $\tilde Q$ and $\dot{\tilde{P}}$.
Even with these constraints, the quantities $\dot Q$ and $\dot{\tilde{P}}$ can be taken as independent, and we get,
\begin{equation}
P=\frac{\partial F}{\partial Q},\qquad \tilde{Q}=(-1)^{a_pa_q}\frac{\partial
F}{\partial\tilde{P}},\qquad H^\prime=H+
\frac{\partial F}{\partial t}+K\label{hj}.
\end{equation}
If, as usual, the new coordinates, $\tilde P=(\tilde
p,\tilde\pi)$ and $\tilde{Q}=(\tilde{q},\tilde{\psi})$, are assumed to be constant, which is guaranteed if $H^\prime=0$ or, what is the same, $H^\prime=K$, then the HJ equation will be in fact given by
a system of equations. If the first equation in (\ref{hj}) is applied
to the last one and to (\ref{constricciones}), we get the HJ equation,
\begin{equation}
H\left [Q,\frac{\partial F}{\partial q}(Q,\tilde P,t),t\right]+\frac{\partial F}{\partial t}(Q,\tilde P,t)=0,\label{hamgras}
\end{equation}
as well as,
\begin{equation}
\frac{\partial F(Q,\tilde P,t)}{\partial \psi_\alpha}=f^\alpha(Q).\label{2classhj}
\end{equation}
Additionally we have the second equation in (\ref{hj}), which can be written as,
\begin{eqnarray}
\frac{\partial F(Q,\tilde P,t)}{\partial\tilde p_i}&=&\tilde q^i={\rm even\ Grassmann\
constant},\label{evengr}\\
\frac{\partial F(Q,\tilde P,t)}{\partial\tilde\pi^\alpha}&=&-\tilde\psi_\alpha={\rm odd\ Grassmann\
constant},\label{grasmans}
\end{eqnarray}
plus the $\mu$ (unknown) constraints (\ref{constr}), which eliminate half of the fermionic constants
$(\tilde\psi,\tilde\pi)$.
Usually, the configuration space variables can be obtained from equations (\ref{evengr}) and
(\ref{grasmans}), as functions of two
integration constants, and this will be the case of (\ref{evengr}), from which the bosonic variables $q$ can be obtained in terms of $\tilde{p}$, $\tilde q$, $\psi$ and $\tilde{\pi}$.
However, before solving the equations (\ref{grasmans}), we can solve the equations (\ref{2classhj}). Indeed,
the fact that (\ref{constr}) are second class means that $f^\alpha$ are invertible, and a solution for $\psi$ in terms of $\tilde p$ and $\tilde\pi$ can be obtained, after substituting $q$ by its solution.
If this solution is then substituted in (\ref{grasmans}), $\mu$ relations among the constants $\tilde\pi$ and
$\tilde\psi$ arise, which will eliminate half of them.

Consistently with these results, we have that, for an action with standard kinetic fermionic term, as a consequence of (\ref{2classhj}) the boundary condition for $F$ will be fulfilled,
\begin{equation}
\delta F(t_1)=\delta\psi_\alpha(t_1)\frac{\partial F}{\partial\psi_\alpha}(t_1)=
\delta\psi_\alpha(t_1)f^\alpha(t_1)=\delta\psi_\alpha(t_2)f^\alpha(t_2)=\delta F(t_2).
\end{equation}

Thus, all these equations (\ref{hamgras}-\ref{grasmans}), must be solved to get the complete solution for
the Hamilton principal function (Hpf),
$$F(Q,\tilde P,t)=S(Q,\tilde P,t)+\alpha,$$
as well as the solution for the configuration space variables $(q,\psi)$, depending on the correct number of integration constants, two for each bosonic degree of freedom, and one for each fermionic degree of freedom.

\section*{A system $L=\psi\dot\psi$}
In this section, a simple instance is solved to show the problems which
appear when fermionic variables are present.

Consider a system characterized by the Lagrangian
$L=\psi\dot\psi$, the Euler-Lagrange equation is $\dot\psi=0$.
The canonical momentum to the fermionic variable $\psi$, is given by
$\pi={\partial L}/{\partial\dot\psi}=-\psi$.
The Hamiltonian is given by
$H_0=\dot{\psi}\pi-L=\dot{\psi}\pi-\psi\dot{\psi}=\dot{\psi}(\pi+\psi)$,
where the velocity $\dot{\psi}$ can be handled as a new parameter. It vanishes, weakly, according to the second class constraint.

Now the HJ formalism is applied, by substituting $\pi=\partial
S/\partial\psi$. In this case the Hamiltonian vanishes and the HJ equation is given by
\begin{equation}
\frac{\partial S}{\partial t}=0,\label{peq}
\end{equation}
where the action depends on the configuration variable $\psi$ and on one constant fermionic parameter $\rho$, i.e. $S=S(\psi,\rho)$. We have as well the equations,
\begin{equation}
\frac{\partial S}{\partial \psi}=-\psi, \qquad
\frac{\partial S}{\partial \rho}=\beta,\label{peq1}
\end{equation}
where $\beta$ is a constant grassman parameter.

Due to the fact that the action is bosonic, it must have the form
\begin{equation}
S=a(t)\rho \psi.
\end{equation}
Applying to it the first equation in (\ref{peq1}), we get $\psi=a\rho$, then we apply the
second equation and $\beta=a\psi=a^2\rho$. Thus $a$ is a constant, as would result
also from (\ref{peq}). Thus, the constant fermion solution of the Euler-Lagrange
equations turns out.

\section*{Interacting system}
In this section, an interacting system with two fermionic variables
$\psi_1$ and $\psi_2$, such that each one are the complex conjugated from
the other ${\psi_1}^*=\psi_2$,
will be discussed,
\begin{equation}
L=i(\psi_1\dot{\psi}_2 + \psi_2\dot{\psi}_1) + k\psi_1\psi_2.\label{interact}
\end{equation}
The Euler-Lagrange equations are given by,
\begin{equation}
i\dot{\psi}_1+\frac{k}{2}\psi_1=0,\qquad
i\dot{\psi}_2-\frac{k}{2}\psi_2=0,\label{acopla}
\end{equation}
with solutions,
\begin{equation}
\psi_1 (t)=\xi_1 e^{(ik/2)t},\qquad
\psi_2 (t)=\xi_2 e^{(-ik/2)t}.\label{s1}
\end{equation}

The Hamiltonian of this system is given by,
\begin{equation}
H=-k\psi_1\psi_2,\label{hj1}
\end{equation}
which must be accompanied by the second class constraints, as definitions of the momenta,
$\pi_1 =- i\psi_2$ and $\pi_2 = - i\psi_1$.
As a consequence, the Hpf will be the
solution of the following system of equations,
\begin{eqnarray}
H(\psi)+\frac{\partial S(\psi,\rho,t)}{\partial t}&=&0,\label{guno}\\
i\psi_2+\frac{\partial S(\psi,\rho,t)}{\partial \psi_1}&=&0,\label{gdos}\\
i\psi_1+\frac{\partial S(\psi,\rho,t)}{\partial \psi_2}&=&0,\label{gtres}\\
\beta_1-\frac{\partial S}{\partial\rho_1}&=&0,\label{gcuatro}\\
\beta_2-\frac{\partial S}{\partial\rho_2}&=&0,\label{gcinco}
\end{eqnarray}
where, $\rho_i=\tilde\pi_i$  and
$\beta_i=\tilde\psi_i$ are constant
odd Grassmann quantities, which satisfy $\rho_1^*=-\rho_2$, $\beta_1^*=\beta_2$, and $\dot\psi_1\pi_1+\dot\psi_2\pi_2$ is real. Seemingly, there are too many constants for
a first order system. However, as will be shown further, the role of the equations (\ref{gcuatro}, \ref{gcinco}) is precisely to establish relations, corresponding to the second class constraints, between them.

In order to solve this system, we write the most general even Grassmann
function of the odd Grassmann quantities $\rho_1$, $\rho_2$, $\psi_1$, $\psi_2$:
\begin{equation}
S(\psi,\rho,t)=S_0(\rho,t)+S_1(\rho,t)\psi_1+
S_2(\rho,t)\psi_2+S_3(\rho,t)\psi_1\psi_2,\label{funcion}
\end{equation}
where the fermionic functions are given by $S_1(\rho,t)=
s_1(t)\rho_1$ and $S_2(\rho,t)= s_2(t)\rho_2$, and the bosonic ones by
$S_0(\rho,t)=s_0(t)+s_{01}(t)\rho_1\rho_2$ and $
S_3(\rho,t)=s_{30}(t)+s_3(t)\rho_1\rho_2$.

From the reality of $S$, we get that the coefficients $s_0$, $s_{01}$,
$s_{30}$ and $s_3$ must be real and $s_1^*=s_2$.

Further we have the conditions (\ref{gdos}, \ref{gtres}, \ref{gcuatro}, \ref{gcinco} )
\begin{eqnarray}
\frac{\partial S}{\partial
\psi_1}&=&-s_1\rho_1+(s_{30}+s_3\rho_1\rho_2) \psi_2= -i\psi_2,\label{pi1}\\
\frac{\partial S}{\partial
\psi_2}&=&-s_2\rho_2-(s_{30}+s_3\rho_1\rho_2) \psi_1= -i\psi_1,\label{pi2}\\
\frac{\partial S}{\partial \rho_1}&=& s_{01}\rho_2+s_1\psi_1+
s_3\rho_2\psi_1\psi_2=\beta_1,\label{pi3}\\
\frac{\partial S}{\partial \rho_2}&=& -s_{01}\rho_1+s_2\psi_2-
s_3\rho_1\psi_1\psi_2=\beta_2.\label{pi4}
\end{eqnarray}

The first two equations, can be solved for $\psi_1$ and $\psi_2$, thus obtaining,
\begin{equation}
\psi_1=-\frac{s_2}{s_{30}-i}\rho_2,\qquad
\psi_2=\frac{s_1}{s_{30}+i}\rho_1,\label{psi1s}
\end{equation}
which substituted in the second two, (\ref{pi3}, \ref{pi4}), give us,
\begin{eqnarray}
\beta_1&=&\left(s_{01}-\frac{s_1 s_2}{s_{30}-i}\right)\rho_2,\label{beta1}\\
\beta_2&=&\left(-s_{01}+\frac{s_1 s_2}{s_{30}+i}\right)\rho_1.\label{beta2}
\end{eqnarray}
These equations could be identified with the second class constraints,
letting only two free constants, as corresponds to fermionic theories.

Taking into account the fact that $\beta_1$, $\beta_2$, $\rho_1$ and $\rho_2$ are
constant, we see that the coefficients in (\ref{beta1}) and (\ref{beta2}) are
themselves constant, that is
\begin{equation}
s_{01}-\frac{s_1 s_2}{s_{30}-i}=A,\qquad
s_{01}-\frac{s_1 s_2}{s_{30}+i}=A^*,
\end{equation}
if we set $A=u-iv$, we get now
\begin{eqnarray}
s_{01}&=&vs_{30}+u,\label{beta11}\\
s_1 s_2&=&v(s_{30}^2+1).\label{beta22}
\end{eqnarray}

Now, in order to write the HJ equation, we note that it has to be written before
substituting (\ref{beta11}) and (\ref{beta22}) into the Hpf, because the time derivative in the HJ equation does not act on the fermionic variables $\psi$. Thus we have,
\begin{equation}
\frac{\partial S}{\partial t}+H=\dot s_0+\dot s_{01}\rho_1\rho_2+ \dot
s_1\rho_
1\psi_1+\dot s_2\rho_2\psi_2+(\dot s_{30}+\dot
s_3\rho_1\rho_2)\psi_1\psi_2-k\psi_1\psi_2=0.
\end{equation}
Taking into account (\ref{psi1s}) we get,
\begin{equation}
\dot s_0+\frac{1}{s_{30}^2+1}\left[\dot s_{01}( s_{30}^2+1)-\dot s_1s_2
(s_{30}+i)-s_1\dot s_2(s_{30}-i)+s_1s_2(\dot
s_{30}+k)\right]\rho_1\rho_2=0,
\end{equation}
or, equivalently $\dot s_0=0$, and
\begin{equation}
( s_{30}^2+1) \dot s_{01}-s_{30}(s_1\dot s_2+s_2\dot s_1)+i(s_1\dot
s_2-s_2\dot s_1)+s_1s_2(\dot s_{30}+k)=0.\label{hj2}
\end{equation}

From the fact that $s_1$ is the complex conjugated of $s_2$ and $ s_{30}$ is
real, and writing in (\ref{beta22})
$v=aa^*$, we get
\begin{equation}
s_1=a^*(s_{30}+i)e^{i\tau},\qquad
s_2=a(s_{30}-i)e^{-i\tau}.
\label{a1}
\end{equation}
These equations, together with (\ref{beta11}), substituted back into (\ref{hj2}), give
$2\dot\tau+k=0$, i.e.
\begin{equation}
\tau=-\frac{k}{2}t+c.
\end{equation}

Thus, if we set $\xi=-a e^{-ic}\rho_2$, we obtain
\begin{eqnarray}
\psi_1&=&\xi e^{\frac{i}{2}kt}\label{solucion1}\\
\psi_2&=&\xi^*e^{-\frac{i}{2}kt},\label{solucion2}
\end{eqnarray}
which coincide with the solutions (\ref{s1}).

Therefore, the Hpf is given by
\begin{equation}
S=s_0+\left(s_{01}-\frac{s_{30}s_1 s_2}{s_{30}^2-1}\right)\rho_1\rho_2=s_0-\frac{u}{aa^*}
\psi_1\psi_2,
\end{equation}
where $s_0$ is constant.

Note that the undetermined functions $s_{30}$ and $s_3$, do not appear neither in the solutions
(\ref{solucion1}, \ref{solucion2}) nor in the Hpf.
This can be understood from the form of the Hpf (\ref{funcion}) and the equations
(\ref{psi1s}), as the terms containing these functions vanish identically.

\subsection*{Canonical transformation point of view}
Let us consider the solution to the equations (\ref{pi3}, \ref{pi4}) for $\psi_1$
and $\psi_2$. We can get first $\psi_1$ from (\ref{pi3}), then we substitute it in
(\ref{pi4}), from which we get,
\begin{eqnarray}
\psi_1&=&\frac{1}{s_1^2}\left(s_1\beta_1-s_{01}s_1\rho_2-\frac{s_3}{s_2}\rho_2\beta_1\beta_2-\frac{s_{01}s_3}{s_2}
\rho_1\rho_2\beta_1\right),\label{psi1}\\
\psi_2&=&\frac{1}{s_2^2}\left(s_2\beta_2+s_{01}s_2\rho_1+\frac{s_3}{s_1}\rho_1
\beta_1\beta_2-\frac{s_{01}s_3}{s_1}
\rho_1\rho_2\beta_2\right).\label{psi2}
\end{eqnarray}
These equations can be written in the form of canonical transformations. In order to see it, taking into account the last observation of the preceding section, let us set
$S_0=0$, and $s_{30}=0$, in this case (\ref{psi1},\ref{psi2}) are given by,
\begin{eqnarray}
\psi_1&=&\frac{\beta_1}{s_1}-\frac{s_3}{s_1^2s_2}\rho_2\beta_1\beta_2,\label{psi1n}\\
\psi_2&=&\frac{\beta_2}{s_2}+\frac{s_3}{s_2^2s_1}\rho_1\beta_1\beta_2 \label{psi2n}.
\end{eqnarray}
Note that these equations are symmetrical under the interchange
$\psi_1\leftrightarrow \psi_2$, $\rho_1\leftrightarrow \rho_2$ and
$\beta_1\leftrightarrow \beta_2$.
If we define $\psi_1^o={s_1^{-1}}{\beta_1}$,
$\psi_2^o={s_2^{-1}}{\beta_2}$, $\pi_1^o=-{s_1}{\rho_1}$,
$\pi_2^o=-{s_2}{\rho_2}$, $\alpha=({s_1s_2})^{-1}s_3$, and the
function $G=\pi_1^o\pi_2^o\psi_1^o\psi_2^o$, equations
(\ref{psi1n},\ref{psi2n}) can be written as
\begin{eqnarray}
\psi_1&=&\psi_1^o+\alpha\frac{\partial{G}}{\partial\pi_1^o},\label{psi1na}\\
\psi_2&=&\psi_2^o+\alpha\frac{\partial{G}}{\partial\pi_2^o}\label{psi2na}.
\end{eqnarray}
Similarly, the momenta (\ref{pi1},\ref{pi2}), can be rewritten as
\begin{eqnarray}
\pi_1&=&\pi_1^o+\alpha\frac{\partial{G}}{\partial\psi_1^o},\label{pi1na}\\
\pi_2&=&\pi_2^o+\alpha\frac{\partial{G}}{\partial\psi_2^o}\label{pi2na}.
\end{eqnarray}
In vectorial notation, these equations can be expressed by a single equation
\begin{equation}
\Delta\bf{u}=\bf{u}-\bf{u^o}=\alpha{J}\frac{\partial{G}}{\partial{\bf{u}}},\label{canonica}
\end{equation}
where $\bf{u}=(\psi_1,\psi_2,\pi_1,\pi_2)$,
$\bf{u^o}=(\psi_1^o,\psi_2^o,\pi_1^o,\pi_2^o)$, and $J$ is the corresponding Jacobian
matrix.

As it can be observed, the function $G$ plays the role of the generating
function of a canonical transformation
(\ref{canonica}), with a finite parameter $\alpha$. The solution needs additional
conditions, for example ``initial conditions'' $\psi_1^o\propto\pi_2^o$
and $\psi_2^o\propto\pi_1^o$. This way to write the solution to the HJ equation, could be useful for computing the Van Vleck determinant for supersymmetric theories \cite{zumino}.

\subsection*{Quantum Mechanical features}
Defining, ${\bf\psi}=(\psi_1, \psi_2)$, ${\bf\psi}^o=(\psi_1^o,
\psi_2^o)$, the equations (\ref{psi1n}, \ref{psi2n}) can be rewritten as follows
\begin{equation}
{\bf\psi}={\bf\psi}^o\exp{(\alpha S)}.\label{wave}
\end{equation}
Indeed, taking $S_0=0$ as in the preceding section,
\begin{equation}
S=-(\pi_1^o\psi_1^o+\pi_2^o\psi_2^o+\alpha\pi_1^o\pi_2^o\psi_1^o\psi_2^o),
\end{equation}
hence
\begin{eqnarray}
\psi_1^o\exp{(\alpha S)}=\psi_1^o+\alpha\pi_2^o\psi_1^o\psi_2^o=\psi_1,\nonumber\\
\psi_2^o\exp{(\alpha S)}=\psi_2^o-\alpha\pi_1^o\psi_1^o\psi_2^o=\psi_2.\nonumber
\end{eqnarray}
Further, considering that
\begin{equation}
\frac{\partial{\bf\psi}^o}{\partial{t}}=(-\frac{\dot{s}_1}{s_1}\psi_1^o,-\frac{\dot{s}_2}{s_2}\psi_2^o),
\end{equation}
it can be seen that ${\bf\psi}$ is a solution of the following first order partial differential equation
\begin{equation}
\frac{1}{\alpha}\frac{\partial{\bf\psi}}{\partial{t}}=\left (\Sigma+\frac{\dot{\alpha}}{\alpha}S-H\right ){\bf\psi},\label{scho}
\end{equation}
where
\begin{equation}
\Sigma=\frac{1}{\alpha}\left (\begin{array}{lr}
-\displaystyle\frac{\dot{s}_1}{s_1}&0\\
0&-\displaystyle\frac{\dot{s}_2}{s_2}\end{array}\right )
\end{equation}
A particular case turns out when the permutational symmetry in eqs. (\ref{psi1n}, \ref{psi2n}) is broken by the application of second class constraints (\ref{psi1s}). If moreover eqs.~({\ref{a1}}) are applied, we get
\begin{equation}
\frac{\dot{s}_1}{s_1}=i\frac{k}{2},\qquad
\frac{\dot{s}_2}{s_2}=-i\frac{k}{2}.
\end{equation}
In this case $\Sigma$ can be written as,
\begin{equation}
\Sigma=\frac{1}{\alpha}\left (\begin{array}{lr}
i\displaystyle\frac{k}{2}&0\\
0&-i\displaystyle\frac{k}{2}\end{array}\right )=i\frac{k}{2\alpha}\left (\begin{array}{lr}
1&0\\
0&-1 \end{array}\right )=i\frac{k}{2\alpha}\sigma_3.
\end{equation}
If for simplicity $s_3$ is assumed to be a constant, eq.~(\ref{scho}) is rewritten as
\begin{equation}
\frac{1}{\alpha}\frac{\partial{\bf\psi}}{\partial{t}}=\left (i\frac{k}{2\alpha}\sigma_3-H\right ){\bf\psi}.\label{scho2}
\end{equation}
This equation resembles the Schr\"odinger equation. Note that, due to the nilpotency of fermionic degrees of freedom, $H{\bf{\psi}}=0$. However, if the Lagrangian (\ref{interact}) would be extended to a supersymmetric one, by the addition of two bosonic degrees of freedom, second order bosonic partial derivatives would appear in eq.~(\ref{scho2}) and the
Schr\"odinger equation of a spinning system of two degrees of freedom would turn out.

Due to eq. (\ref{wave}), we can make the identification $\alpha\equiv{1}/{\hbar}$. Hence, taking into account
({\ref{a1}}), we have the following relation
\begin{equation}
\hbar=\frac{|{a}|^2}{s_3}.
\end{equation}

\begin{acknowledgments}
This work was supported in part by CONACyT M\'exico Grant No. 37851E. P. R. thanks the Department of Applied Physics of Cinvestav-M\'erida, where part of this work was done.
\end{acknowledgments}

\end{document}